# Analysis of Personalized Information Service System for Digital Libraries

Li Qian, Lizhen Liu, Yue Hu
*Capital Normal University, Beijing 100037, P.R.China*
qianli1981119@163.com

**Abstract**

*Along with the rapid development of digital library, digital library of the third generation, taking the main characteristics of the personalized service, has already become the mainstream today. This article first analyzes the concept and characteristics of digital library, then elaborates the inevitability of personalized service as well as the necessity of its development and based on the Propose of establishing interested knowledge-base, enumerates several ways of personalized service, at last according to individualized service, this paper proposes the model of "learning-oriented individual digital library" and prospects the trends of digital library.*

## 1. Introduction

With the rapid development of advanced technology, the internet which represents the technology of the computer and the web communication is growing up extremely. As a result, the human being entered into the information society. Meanwhile, after the traditional period focusing on the MARC managing system of books categories and the automatic period centering on the digital information sources, the library comes into the third generation—digital library, which takes the characteristics of distributed, isomerous, and personalized. In accordance with these characteristics and the surge of digital information sources, digital library must carry out the guidance of user-central and humanist, with the purpose of supplying individualized information for users and saving the time of they finding information Therefore, in theory supplying personalized service for user has become a key point in the development of digital library, and has partly applied in practice. We analyze profoundly the individualized service of digital library, and effectively organizing the individualized information so as to achieve an initiative, reciprocal, individualized service is one of problems which we should solve first.

## 2. General introduction to digital library

### 2.1. The concept of digital library

In our daily lives, we have realized that the technical conditions and environment of digital library has being improved and perfected gradually, but it is always difficult to define the digital library. At the beginning the concept of digital library is offered by some physical scientists, not librarians, when they are making software for an academic communication on the internet. After the creation of the world-wide-web technology, various information sources and related software comes into being [1]. Now, there are following several typical definitions [2]:

1) American plays a prior role in the construction of digital library as its development. With the implements of its DLI-1 and DLI-2, the concept of digital library in a report of a "State Challenge" item relating to digital library and sponsored by the National Scientific Fund Commission of America is that it is a series of information sources and technical methods organizing these sources (such as the technology of establishing, searching and taking use of the information: It covers all the storage and searching system of digital media styles ( Text, picture, voice, dynamic picture and so on ).

2) Digital library is a system that provides the users with enormous and organized information and knowledge-base that easily search and use.

3) Digital library is a large global and distributed knowledge repository, which is supported by numerous data-based and based on intellectual technology.

4) Digital library is a digital system of information source adopted by modern high-tech, which is a large scale knowledge-center and convenient for use beyond the time and space limits [3].

Nowadays, digital library has experienced its second developing period with the implement of DLI-1 innovate plan, and came into the third period with the driving of DLI-2 innovate plan, which important construction from the interest of their customers. In conclusion, digital library not only makes the traditional library digital and makes the digital source cyberized, but it is a information flat serving for social group; in some extent, it is a kind of service idea for single customer and customer groups.

### 2.2. The characteristics of digital library

On the basis of understanding some definitions of the digital library, it is essential for the development of the digital library to clearly understand its characteristics, because it determines the realization of related technology and the quality of the individualized service for the customs.

With the construction and development of digital library, as the digital library that serves people, it firstly should have these following characteristics: information resource digitization, gaining information resource cyberization, using information resource sharing information resource hypothesized. Besides these, there are others as follows [4]:

1) Metadata: In the earliest research of digital library, American NASA put forward the questions: How to make the catalogue, classification and subject of the material source like librarians in order to make user look up easily. Whereas the traditional library has made the catalogue superficially, such as the list, name, the abstract. So, one of the main features is metadata. Now the scope of the metadata has developed from early describing the characteristics basing on the contents to the metadata system of management, techniques, preservation and description.

2) Distribution: There are several ways in managing data, current development trends to the characteristics of distributed management.

3) Dispatchment: In the systematic construction's research on digital library, the NASA dispatched the distributed sources and objective sources in different units by analytical location, while the use of sources between two traditional libraries depended on their contracts, such as Z39.2 agreement. If there is no contract between them, they can't use reciprocal sources even they have an agreement. Taking use of technology is the key of dispatching system in the digital library.

4) Diversification of Service: Metadata search, objective source service, full text search, diversified guidance, gateway, crossing storehouse, multimedia, virtual reference and consultation, literature transmission, reciprocal loan among library, individuation, delivery of information, reservation service, note service, E-mail service, single point base, systematic base and so on.

5) Standard and criterion: Standards of metadata and object base, criterions of self-establishing sources, abiding extent of industrial, internal, international criterion and so on[5].

## 3. Analysis of personalized service

### 3.1. The definition and the significance of personalized service

**3.1.1. The definition.** It dated back to the new idea put forward by economic domain. After having experienced the time of planned economy, self-center theory has been submerged by today's network, and popularized service is also covered by the requirement of personalized service.

Compared with the popularized service, personalized service is a pattern that aims at supplying different service strategies, contents and functions for different customer. It requires that: first, every customer has its own demands of knowledge; second, as a supplier of the service, we should have a definite object and get effect instantly. At last, the supplier and the receiver of the service should cooperate with each other and make the two party satisfactory commonly.

**3.1.2. The practical significance.** The idea of personalized service takes the user as the center and advocates the thought of taking persons as the foremost, therefore it impacts much significance to the construction of digital library of the third generation. Here summed in four areas:

1) Sustainable development of digital library calls for personalized service, and it even trends to the complex development of both the traditional library and digital library.

2) From the angle of customs, digital library change a library facing to the public into the individual. On the one hand, it is convenient and timesaving for the users, on the other hand, it help to find and cultivate individuation, guide the requirement, and promote the social development of variety.

3) Taking the customer as the center, digital library optimizes the distribution of library's information, human and non-human sources, applies the idea of knowledge economy to the establishment of the library, exerts its function, and advances its quality and efficiency.

4) Personalized service is one of the main characteristics of its third generation.

### 3.2. The key of implementing the personalized service of digital library—the establishment of the interested knowledge –base

**3.2.1. Introduction to knowledge.** In the information science, it is a fact that according to the agreements the data use, the information endows the meaning of the data, and after processing and reestablishment the information, it changes into the knowledge. Based on the human practice and proved by the time, knowledge credibly reflects the impersonal reality. Knowledge is the innovative fruit of our brains and the result of human intelligence. The intelligence is the source of human civilization, the power to history development and the core of many factors of productivity [6].

**3.2.2. The necessity of establishing the interested knowledge-base.** All the people in their fields may easily find the information source what they want in the digital library which works as the center of information sources. Meanwhile the people in this group is relatively stable, so classifying all the people by their interests is in favor of the construction of information source and obtainment of the customer want. In the aspect of establishing information source, in the digital library's development, it is an important part for the development of the digital library to classify information effectively, because we can not take full advantage of digital library without well chain of information source. In aspect of using information source, every customer groups is an entity with the same interest, so we can sort these entities, give relative analysis and at last build a kind of model by investigating data. In conclusion of the above aspects, it is necessary to build interest knowledge-base for these steady groups.

**3.2.3. Superiority of establishment the interest knowledge-based.** In conclusion of the introduction in 3.1, it is an important guarantee for the sustainable development of human intelligence to discovery the knowledge, in turn making use of the knowledge can accelerate the discovery of knowledge, and therefore they both can give a complement each other to the development of social science and technology. It is an essential part for the main construction of present digital library to make it individualized, moreover, it is also the essential part to build the interest knowledge-base in the implement of service with individuality. A nice knowledge chain, which founds customer's interested knowledge-base by knowledge discovery and in turn serves customer, undoubtedly plays a key role in successfully applying personalized service of digital library.

Interested knowledge-base supplies the information sources of customer and suited information for personalized service. The data-mining technology, which is found from the basic information and the behavior track of customer, is a knowledge-base based on the base-mining and combined information sources with the coordination and analysis of information source, its aim is to offer personalized service to user and save their time of judging information. It may also greatly economize manpower and physical resource.

**3.2.4. The establishment of interested knowledge – base.** It is used for keeping customer's information and its interest data from automatic system. Users can obtain and feed back information from exploring the web and system recommendation. The general procedure of establishment is as follows:

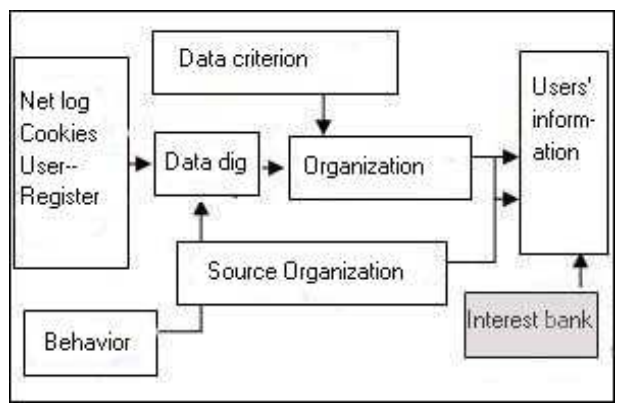

**Figure 1. The process of interested knowledge bank**

### 3.3. The way of personalized service

Accompanied with the process of creative project for the digital library launched by American Natural Science Fund Association, at present DLI-2 clearly proposes one of the most important point for this project, that is the research of person-oriented, which promotes their ability in creating 、 inquiring and making use of the knowledge information and also promotes relative technology in establishing digital library. The way of personalized service will play an important role in this process.

**3.3.1. Personalized digital library.** The development of resource in digital library, and along with the mature development in relative technology of digital library has led to its own massive systematic structure as well as lots of abundant information for users. Because the users' demand for information is limited and their interest is very stable. "personalized digital library" closely revolves the essence that "the user can organize and make use of this source effectively and efficiently". It also embodies the service idea of digital library's person-oriented. As for users, they can freely build their own files according to their own will, upload their special local source, combining with their interested information bank exercise personalized retrieval and order personalized service, which makes them easily share an individual source environment. All this advantages mentioned above for users are absolutely easy---the only thing for them to do is to apply for a space in this public information platform. So we can say the establishment of personalized digital library to some extent on will lessen the pressures from the public digital library, on the other hand owing to wide range of knowledge, it will be more convenient for users.

**3.3.2. Individual retrieval.** The purpose of digital library is to satisfy user's information requirements and it is just the digital library that will provide more convenience for seeking the information. However with the dramatic increase of digit, the potential information source also increases, which brings people much trouble of inquiring for information. The increase of abundant information has wasted their judging time, which will be solved by the individual retrieval. Based on the retrieval original data and combined with user's interested bank, individual retrieval lists their search requirements by which it will present the most proper information before users.

**3.3.3. The personalized recommended service.** This service, which is an embodiment of immediate personalized service, takes the user's interested knowledge-base as the leadership, bases on the latest data source and initiatively offers personalized information service by means of information technology.

The widely-used service in digital library is more closely to the learning environment for all users. Meanwhile it has made use of the high-tech cyber technology, which can provide the latest information in the first time to the user who is seeking them badly. It includes the latest information recommendation about document, special conferences information, the articles in the excellent journals and hot issues etc. We may refer to the following service:

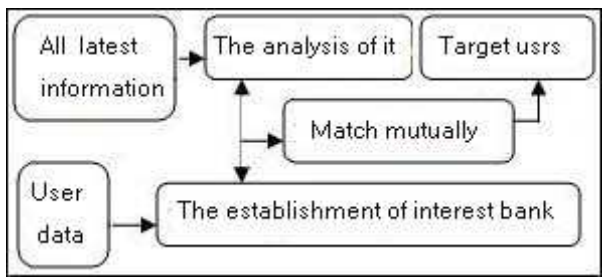

**Figure 2. The flow of the recommended service with individuality**

## 4. Shape the model of learning-oriented digital library

### 4.1. The proposal of learning-oriented digital library

With the development of quality education in this information society, learning has become a life idea of people and their social value also has changed from the passive adoption to initiative learning, which enhances them deeply to learn so as to seek their badly-needed knowledge more conveniently and quickly. Meanwhile present massive higher educational is requiring more efficient knowledge output system by which people can get the needful knowledge in the first time. Faced with this situation, the model to shape the digital library was put forward in order that library can become an open class for people's lifelong education and that they can gain the knowledge at any time and at any place by the advanced computer and cyber technology. Along with the use of service with individuality, the digital library's service efficiency and quality also have been promoted.

### 4.2. The analysis of the learning- oriented digital library

The learning-oriented digital library firstly embodies the concept of knowledge economy basing on the production 、 exchange and use of knowledge and information and all of these have been considered to be the new increasing point of economy in China. As a treasury of knowledge, digital library must conform to the requirement of knowledge economy so as to offer everyone heir interesting knowledge timely 、 precisely and effectively as more as it can. Secondly it takes advantage of the advanced cyber technology, integrates interesting knowledge bank and provides precise knowledge guidance to save their corresponding finding time. Lastly future library is the combination of digital library and traditional library, that is to say it is the complex library. Facing to so complicated situation, offering personalized service will become the first priority, because it is the best guidance for user to dive in the knowledge.

### 4.3. The structure of learning-oriented digital library

The learning-oriented digital library's absolute advantage in utilizing data source is having established a visual personalized learning environment and provides knowledge service for students by Web way. According to their own situation, students can refer to the teaching program conveniently, arrange their time flexible and select any Web place to learn by themselves. By this way in this visual teaching environment, teachers and students can exercise the mutual communication across the space.

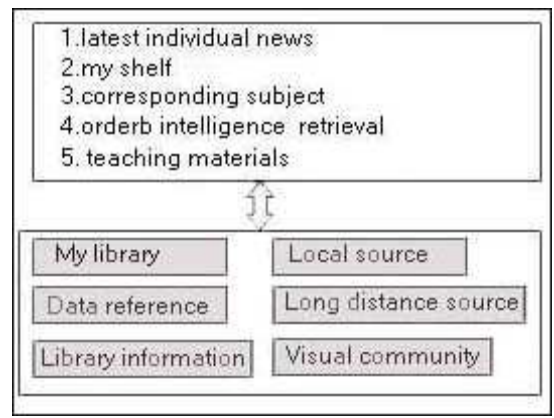

**Figure 3. The structure of the digital library**

## 5. Trends of the digital library

At present, the digital library has become a new commanding elevation in the international high-tech competition and one of the important symbols in the assessment of a country's information infrastructure. The establishment 、development and application of it are doom to affect the quality of all the social life. The digital library as well as the social sustainable development to much extent determines the key junction in science. So it is extremely important for the right prediction in the trend of digital library.

According to the present situation both abroad and at home, together with the development history of digital library, the service-oriented pattern [7] will become the main trend in future. Under the guide of person-oriented and users-oriented, this pattern greatly attentions people's feeling about digital library, its design comes around users and its research mainly revolves around the relations and surface between computer and users, this pattern stresses personalized service and reduces the inference in the course of using digital library. Another important aspect is that under the guidance of the service-oriented pattern and based on digital source, it will solve the abnormal structure in digital library source on the uniform visit platform so as to provide users different mutual operation in different digital library.

## Acknowledgement

This work was supported by the Beijing Educational Committee Science and Technology Development Planned Project：(KM200610028014) and by the Youth teacher fund of Huoying Dong educational foundation：(91101).